\documentclass[aps,prl,twocolumn,groupedaddress]{revtex4}

\usepackage[final]{graphicx}

\begin{document}


\title{Complete identification of alkali sites in ion conducting
       lithium silicate glasses: a computer study of ion dynamics}


\author{Heiko Lammert}
\email[]{hlammert@uni-muenster.de}
\author{Magnus Kunow}
\email[]{kunow@uni-muenster.de}
\author{Andreas Heuer}
\email[]{andheuer@uni-muenster.de}
\affiliation{ Institute of Physical Chemistry, Schlossplatz 4/7, D-48149 M\protect\"unster\\
              and Sonderforschungsbereich 458 }


\date{\today}

\begin{abstract}
The available sites for ions in a typical disordered ionic
conductor are determined.  For this purpose we devised a
straightforward algorithm which via cluster analysis identifies
these sites from long time ionic trajectories below the glass
transition. This is exemplified for a lithium silicate glass $\mathrm{
(Li_2O)}_x\mathrm{(SiO_2)}_{(1-x)}$ for $x = 0.5$ and $x = 0.1$.
We find for both concentrations that the number of sites is only slightly
bigger than the number of ions. This result suggests a
theoretical description of the dynamics in terms of mobile
vacancies as most appropriate. Furthermore identification of the
ionic sites allows one to obtain detailed characteristics of the
ionic motion, e.g. quantification of correlated forward-backward
jumps.
\end{abstract}


\maketitle

Ion conducting glasses have been investigated by various
experimental methods, including EXAFS
\cite{greaves:1991,rocca:1992}, NMR
\cite{maekawa:1991,yap:1995,gee:1996},
and conductivity spectroscopy
\cite{funke:1993,abe:1993}. Whereas quite
detailed information about the local structure has become
available the mechanism of dynamics is still under debate although
consensus has been reached that ion dynamics can be described as
jumps of the mobile ions \cite{ingram:1987,baranowskii:1999,bunde:1994}.
Whereas some authors stress the relevance of the disordered energy
landscape \cite{baranowskii:1999,dyre:2000} supplied by the network, others relate
the complexity of ion dynamics to the Coulomb interaction among mobile
ions \cite{funke:1993}. Furthermore it has been argued from structural
considerations that the distribution of alkali ions is inhomogeneous
\cite{greaves:1985}.

For a closer understanding of ion dynamics microscopic information as
supplied by molecular dynamics (MD) simulations is highly welcome. Jund
et al. \cite{jund:2001} found preferential pathways as a dynamical
phenomenon by counting the number of different alkali ions that
passed through subvolumes of the simulation box. The resulting
subvolumes visited by the largest number of different alkali ions
may be interpreted as fast areas. They form a network of
conduction paths. Oviedo and Sanz have argued on a qualitative
basis that for alkali concentrations lower than 10\% the alkali
ions are always surrounded by non-bridging oxygens (NBOs) despite
the overall small number of NBOs \cite{oviedo:1998}. This implies
that new NBOs are formed via breaking of Si-O bonds along the
alkali trajectory. In contrast, for higher concentrations hopping
dynamics between so-called micro-channels is proposed such that no
formation of new NBOs would be necessary. Horbach \cite{horbach:2002}
showed in simulations of ${\mathrm{(Na_2O)}2\mathrm{(SiO_2)}}$,
that the coherent intermediate scattering function of sodium
relaxes only on the same timescale as those of the network species,
proving the existence of stable alkali sites below the glass
transition.
Cormack \cite{cormack:2002} et al. have recently
investigated the mechanism of sodium migration in simulations of
${\mathrm{(Na_2O)}_{0.25} \mathrm{(SiO_2)}_{0.75}}$ glasses, observing a
few sequences of jumps between selected sites. He interpreted the
resulting dynamics as the motion of vacancies and pointed out that
the identification of \textit{all} sites in the glass, empty or
populated at a given time, would be most useful for obtaining a
deeper understanding of the kinetic mechanisms.

Determination of all alkali sites in the silicate network is the
purpose of this work. These sites can be defined as any regions in
the network where alkali ions stay. They must correspond to
local minima of the potential energy for the ions.
Thus determination of the ionic sites is equivalent to
determining the effective energy landscape experienced by the
ions. The potential energy involves the network contribution as
well as the averaged interaction with other ions.

We analyse lithium silicate glasses $\mathrm{
(Li_2O)}_x\mathrm{(SiO_2)}_{(1-x)}$ with $x = 0.5 $ and $x = 0.1$ which are
common model systems for computer investigations of ion dynamics.
The system size has been chosen such that for $x=0.5$ we have 384
lithium ions and for $x=0.1$  80 lithium ions. Simulations are
performed in the NVT ensemble at temperatures $640\,{\textrm K}$ and
$920\,{\textrm K}$ and simulation box lengths $22.96$ \AA{} and $25.28$
\AA{}, respectively. The resulting densities match experimental values
\cite{doweidar:1996}. The temperatures have been chosen such that the
lithium diffusion constants in both systems have similar values.
We use a pair potential developed by J.
Habasaki \cite{habasaki:1992}. Previous studies with this
potential have shown good agreement with experimental results for
static and dynamic quantities
\cite{habasaki:1995,banhatti:2001,heuer:2002}. The
molecular dynamics simulations are performed with a modified
version of Moldy \cite{refson:2000}. Since both temperatures are
below the respective glass transition temperatures we proceed in
two steps. We start from a configuration which was first
equilibrated just above the glass transition. It is then
propagated at the selected temperature for ca. 20 ns in the NVT
ensemble until the lithium subsystem is equilibrated. Afterwards
we perform a production run for approximately 50 ns. During this run,
positions of all particles are recorded every 0.1 ps. More details about
the simulations can be found in \cite{heuer:2002}. We checked the
validity of our results for independent data of up to 10 ns duration.
This time is long enough for the lithium dynamics to become diffusive, 
so that just the dynamics relevant for ionic transport is included.

\begin{figure}
\includegraphics[width=8.6cm]{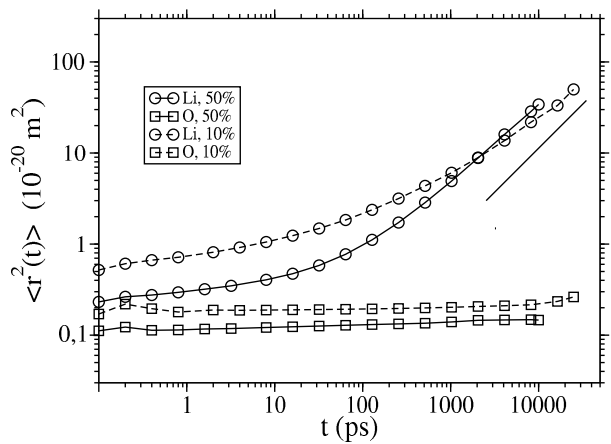}
\caption{\label{fig_msds}Mean square displacements of lithium and
oxygen in the simulations. Included is a line with slope 1.}
\end{figure}

In Fig. \ref{fig_msds} we report the mean square displacement of
oxygen and lithium. The dynamics of the lithium ions becomes diffusive
for $t_{diff} \approx 1$ns for $x = 0.5$ and $t_{diff} \approx 10$ ns
for $x = 0.1$, as obtained from longer simulations.
Evidently, the oxygen ions are localised on
the ns-time scale.  The network of silicon and oxygen atoms and
with this the lithium sites can therefore be considered as static.
As a consequence the ionic sites are basically unmodified
during our run. Interestingly, as shown recently \cite{Sunyer:2002,angell:1981}
the remaining local fluctuations of the network are nevertheless
essential in promoting the ionic jumps.

Based on the lithium trajectories we have devised a
straightforward algorithm to identify ionic sites. As a basic idea
we let the ions decide via their dynamics where good sites
are rather than determining the potential energy, as done e.g. in
\cite{habasaki:1996}. First the simulation box is divided into
cubic cells with size approx. (0.3 \AA)$^3$. They are small enough
to enable resolution of the shape of the ionic sites. During
our MD-run we count how many time steps a cell is visited by
a lithium ion. The cells with nonzero counts describe the portion
of the system that has been visited by lithium ions. These cells
include the lithium sites as well as the connecting paths between
them. To identify the ionic sites and to eliminate the paths
between the sites, cells with less than a minimum value $M$ of counts
are dismissed. With the remaining cells a cluster analysis is performed.
Cells which share a face are grouped into one cluster. Cells that
are not directly or indirectly connected over common faces form
different clusters. The value of $M$ has been determined by the
condition that the number of distinct ionic clusters be a maximum.
For a smaller threshold some lower populated clusters would not
have been detected, for a larger threshold distinct clusters would
be connected to single clusters because the transition path would
be included. With $M$ chosen according to this condition, 2.9\% and 1.3\%
of all cells are included for $x = 0.5$ and $x = 0.1$, which contain
$\approx 90$\% of all lithium positions in both cases.
The precise value of $M$ is of no relevance for the latter analysis.

\begin{figure}
\includegraphics[width=7.0cm]{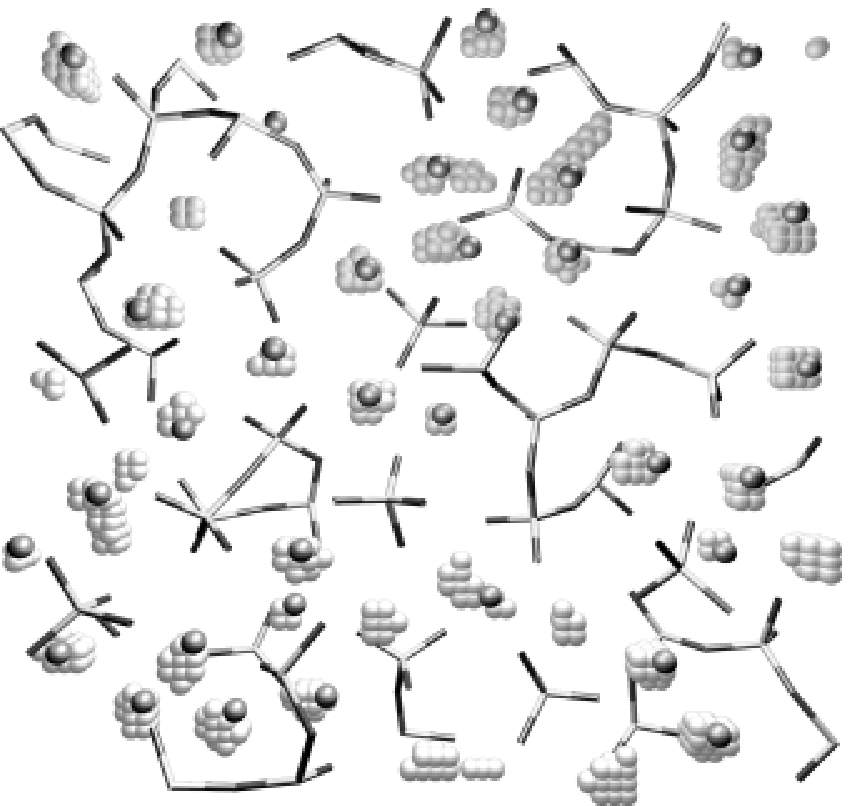}

\includegraphics[width=7.0cm]{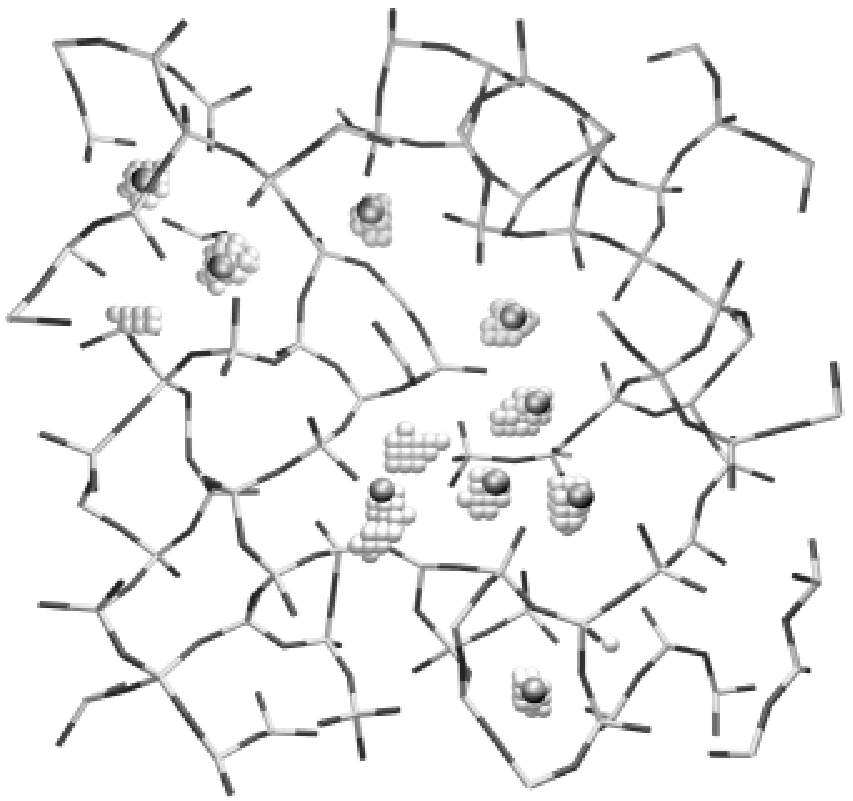}
\caption{\label{fig_slice}Rendering of a slice of the simulation box for
$\mathrm{(Li_2O)}_x\mathrm{(SiO_2)}_{1-x}$ with $x=0.5$(top) and $x=0.1$(bottom),
together with clusters of cells with high average lithium occupation. (Created with the software vmd \cite{vmd}.) }
\end{figure}

For $x=0.5$ and $x=0.1$ respectively, 378 and 76 clusters are found.
The result is illustrated in Fig.
\ref{fig_slice}, showing a snapshot of a slice through each
system. The silicate network is shown as tubes, with silicon
bright and oxygens dark. Lithium ions are depicted as large
spheres. The white objects are the clusters obtained from the
analysis described above, with each small white sphere
representing one of the cells. It can be seen that the clusters
are in fact discrete objects of compact shape. Most of the lithium
ions reside inside of one of these clusters, and most of the
clusters are occupied by exactly one lithium ion. As will be
quantified further below some clusters can accommodate up to three
lithium ions. 

\begin{figure}
\includegraphics[width=8.6cm]{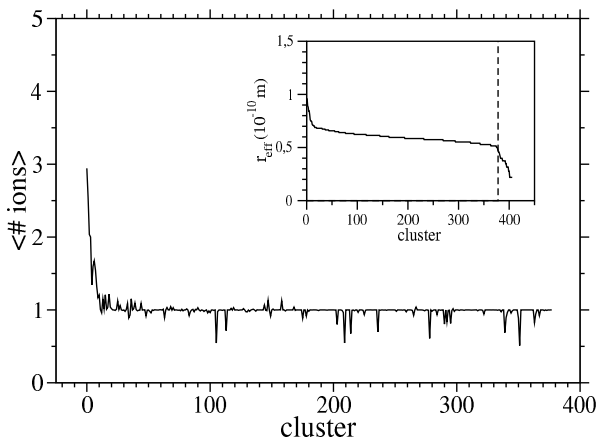}
\caption{\label{fig_occ} Mean occupation of clusters. Inset: effective radius
$r_{eff} \equiv (3V/4\pi)^{1/3}$ of clusters, including the small
clusters excluded from further analysis (see text). }
\end{figure}

For x=0.5 (x=0.1) in total 29 (14) additional small clusters
have been excluded. They are set apart by a conspicious drop in
the effective radius plot; see Fig. \ref{fig_occ}. Detailed
analysis characterised most of them as simple satellites of bigger clusters.
The remaining ones serve as short--lived
saddle--like states during transitions between bigger clusters.
The small clusters need thus not be considered as sites themselves.

On the basis of the clusters, the dynamics of a lithium ion can
be expressed as successive residences in clusters separated by
shorter time intervals where the lithium ion is outside of any
cluster. When a lithium ion leaves one cluster and moves into a
different one, this sequence of events is recorded as a jump. In
contrast, if an ion explores part of the volume outside a cluster
without entering a second one and comes back to the original
cluster, it is regarded as occupying the same cluster during the
whole time. Consequently, the residence time of a lithium ion in a
cluster is defined as the time between its jump into this cluster
and the subsequent jump to another cluster. With this definition
on average a lithium ion in the system with $x=0.5$ belongs to a
cluster at 99.5\% of all times. The remaining 0.5\% correspond to
jumps between different sites. For $x=0.1$, the portion of times
corresponding to jumps is 1.1\%.
The typical distance travelled by the lithium ions during a jump
coincides well with the first maximum of the radial distribution
function of the clusters in both cases.
For $x=0.5$, 93\% of all jumps are inside the first neighbour shell
(94\% for $x=0.1$) as defined by the first minimum of the radial
distribution function. Thus the ion dynamics can be described as jumps
between adjacent clusters.

Having defined the residence times the mean number of lithium ions
occupying each cluster can easily be evaluated.  It is plotted in
Fig. \ref{fig_occ} for $x=0.5$, with the clusters ordered from
left to right by decreasing volume $V$ (the effective radius
$r_{eff} \equiv (3V/4\pi)^{1/3}$ of the individual clusters is
plotted in the inset). The occupation data confirms the impression
gained from Fig. \ref{fig_slice}. For most clusters the value is
close to unity. These clusters are occupied by a single ion during
the largest part of the simulation. Their effective
radius $r_{eff} \approx 0.6$\AA{} is very similar. Only the
largest clusters offer space for more than one lithium ion, i.e.
they contain more than one site, and often they are indeed
multiply occupied.

Generally an empty site must be available for a jump of a lithium
ion. The total number of sites can now be estimated from the
occupation data in Fig. \ref{fig_occ}. For the larger clusters
with cluster index $ <19 $ we have taken the next higher integer
number to estimate the number of available sites. In this way we
find 3 clusters with 3 sites and 9 clusters with 2 sites.
Starting from cluster index 14 one finds a few clusters with
occupation number 1 + $\epsilon$ with $\epsilon < 0.2$. Here it
would be rather unphysical to attribute two sites to this cluster.
Rather a more detailed analysis of explicit ion trajectories has
revealed the following scenario. At most time instances a single
ion stayed in this cluster. During the short times during which
this ion explored the immediate neighborhood of this cluster
without entering a new cluster (typical excursion length scales
are 2 \AA{} from the center of the cluster) a second ion may
briefly enter this cluster and immediately leave this cluster
again. Thus one may either say that these clusters only contain a
single site or, to be more conservative with respect to finding an
upper bound of the total number of sites, attribute $1+\epsilon$
sites to this cluster. With the latter variant we obtain 395 sites
for our 384 lithium ions. Analogously, 86 sites were found for the
80 ions for $x=0.1$. These results shed new light on the mechanism
of ion dynamics. Rather than speaking of individual ions jumping
between accessible sites it is more appropriate to speak to first
approximation of vacancy dynamics, i.e. the dynamics of the
non--interacting unoccupied sites in the energy landscape.

Actually, Dyre \cite{Dyre:2002} has recently suggested this type
of scenario based on general reasoning. He relates the possible
near-equality of sites and ions to the effect that during network
formation at the glass transition only a minimum number of sites
will be formed due to energetic reasons. This scenario is also
compatible with the counter-ion model, proposed by Dieterich and
coworkers \cite{Dieterich:1994}, but is at variance with
single-particle approaches.

\begin{figure}
\includegraphics[width=8.6cm]{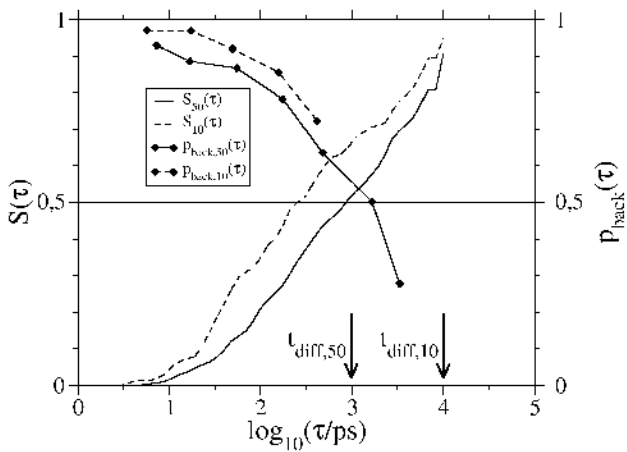}
\caption{\label{fig_times}Cumulative distribution function $S(\tau)$
of residence times and time dependence of the backjump probability
$p_{back}(\tau)$ for systems with $x=0.5$ and $x=0.1$. Arrows mark
the times at which the diffusive regime is reached.
}
\end{figure}

In the remaining part of this paper we would like to characterize
the ion dynamics in terms of the individual clusters. Fig.
\ref{fig_times} shows the cumulative distribution function $S(\tau)$
of residence times. Thus $S(\tau)$ denotes the fraction of clusters
with an average residence time less than $\tau$. Its median
$\tau_{median}$, defined by $S(\tau_{median}) = 0.5$,  is close to 1 ns
for $x=0.5$ and 250 ps for $x=0.1$. These numbers have to be
compared with the median of the transition time for a jump between
two clusters, which is 500 fs for $x=0.5$ and 200 fs for $x=0.1$.
This dramatic time scale separation clearly justifies the hopping
picture in this kind of materials, as known from previous simulations
\cite{habasaki:1997}. The lithium dynamics on long time scales can
thus be mapped to jumps between the clusters.
The generally higher frequency and lower duration of jumps at $x=0.1$
can be traced to the higher temperature that was chosen for this system to
achieve similar diffusivities.

The effectiveness of jumps for diffusive transport is strongly decreased
by correlated forward-backward motion. We determined the probability
$p_{back}$ that a jump from a cluster A to another cluster B is followed
by a direct backjump to the cluster A. The result is also shown in
Fig. \ref{fig_times}. One can clearly see that clusters with short residence
times show a stronger tendency for a correlated forward-backward jump.
For the fastest clusters, $p_{back}$ reaches more than 0.9, which is several
times higher than the statistical value given by the inverse number of nearest
neighbors $(\approx 0.2\ldots{}0.3)$.
The difference in $p_{back}(\tau_{median})$, which is $\approx 0.8$ for
$x=0.1$ compared to $\approx 0.55$ for $x=0.5$, may explain why the
former system becomes diffusive only after most ions have performed
several jumps, as shown by the ratio $\tau_{median}/t_{diff} >> 1$.
In contrast one finds $\tau_{median}/t_{diff} \approx 1$ for $x=0.5$,
meaning that only half of the ions have jumped at least once at the
onset of diffusion. The dispersive regime at shorter times is thus
dominated by the fastest lithium ions.

Beyond the results shown in Fig. \ref{fig_times}, a detailed
analysis of the dynamics, i. e. in terms of cooperativity,
becomes feasible.
Our approach may easily be
generalised to more complex systems 
and important questions concerning, e.g., the character and the
lifetime of the individual sites may be analysed.


We acknowledge interesting discussions with B. Doliwa, J. Dyre, M. Ingram,
P. Jund, B. Roling, and M. Vogel and very helpful correspondence with
J. Habasaki about this topic.


\end{document}